\documentclass{article}
\usepackage{amssymb}
\def\ds{\displaystyle}
\def\g{\gamma}

\def\res{\mathop{\mathrm {res}}\limits_}

\makeatletter
\@addtoreset{equation}{section}

\newtheorem{theorem}{Theorem}[section]
\newtheorem{examp}{Example}[section]
\newtheorem{coroll}{Corollary}[section]
\newtheorem{examps}{Examples}[section]

\newtheorem{lemma}{Lemma}[section]
\newtheorem{remark}{Remark}[section]

\newtheorem{remarks}[remark]{Remarks}
\newtheorem{proposition}{Proposition}[section]
\newtheorem{definition}{Definition}[section]

\def\le{\left}

 \def\tr{{\rm Tr}}

\def\ri{\right}
\def\br{\begin{remark}\rm\small}
\def\1{{\bf 1}}
\def\er{\end{remark}}
\def\bt{\begin{theorem}\rm}
\def\et{\end{theorem}}
\def\bc{\begin{coroll}\rm}
\def\ec{\end{coroll}}
\def\brs{\begin{remarks}.\\ \rm\small\begin{enumerate}}
\def\ers{\end{enumerate}\end{remarks}}
\def\bx{\begin{examp}\small}
\def\ex{\end{examp}}
\def\bl{\begin{lemma}\small}
\def\el{\end{lemma}}
\def\bxs{\begin{examps}. \rm\begin{enumerate}}
\def\exs{\end{enumerate}\end{examps}}
\def\bd{\begin{definition}}
\def\ed{\end{definition}}
\def\bp{\begin{proposition}\rm}
\def\ep{\end{proposition}}
\def\be{\begin{equation}}
\def\ee{\end{equation}}

\def\bea{\begin{eqnarray}}
\def\eea{\end{eqnarray}}
\def\beas{\begin{eqnarray*}}
\def\eeas{\end{eqnarray*}}
\def \pa{\partial}
\def\C{{\mathbb C}}

\def\R{{\mathbb R}}
\def\N{{\mathbb N}}
\def\P{{\cal P}}

\def\a{{\alpha}}
\def\b{{\beta}}
\begin{document}
\begin{flushright}
CRM-2921 (2003)\\
\end{flushright}
\vspace{0.2cm}
\begin{center}
\begin{Large}
\textbf{Free Energy of the Two--Matrix Model/dToda Tau--Function}
\end{Large}\\
\vspace{1.0cm}
\begin{large} {M.
Bertola}$^{\dagger\ddagger}$\footnote{bertola@mathstat.concordia.ca}
\end{large}
\\
\bigskip
\begin{small}
$^{\dagger}$ {\em Department of Mathematics and
Statistics, Concordia University\\ 7141 Sherbrooke W., Montr\'eal, Qu\'ebec,
Canada H4B 1R6} \\ 
\smallskip
$^{\ddagger}$ {\em Centre de recherches math\'ematiques,
Universit\'e de Montr\'eal\\ C.~P.~6128, succ. centre ville, Montr\'eal,
Qu\'ebec, Canada H3C 3J7} \\
\end{small}
\bigskip
\bigskip
{\bf Abstract}
\end{center}
\begin{center}
\begin{small}
\parbox{13cm}{
We provide an integral formula for  the free energy of the two-matrix model with polynomial
potentials of arbitrary degree (or formal power series). This is known
to coincide with the $\tau$-function of the dispersionless
two--dimensional Toda hierarchy. The formula generalizes the case of  conformal maps of Jordan
curves studied by Kostov, Krichever, Mineev-Weinstein, Wiegmann, Zabrodin and
separately Takhtajan. Finally we generalize the formula found in genus zero to
 the case of spectral curves of arbitrary genus with certain fixed data.}
\end{small}
\end{center}
%
%
%
%
%
%
%
%
%
\section{Introduction}
Many instances of integrable systems are obtained by means of a
suitable limit (dispersionless or semi-classical) of a statistical
theory. The departing point of our analysis in this paper is the
random 2-matrix model \cite{Mehta, BI}, which is attracting growing attention due to
its applications to  solid state
physics \cite{Guhr} (e.g., conduction in mesoscopic devices,  quantum chaos and,
lately, crystal growth\cite{spohn}), particle  physics \cite{Verbaarshot},
$2d$-quantum gravity and string theory \cite{Matrixsurf, ZJDFG, DVV}.
The model under inspection consists of two Hermitian matrices
$M_1,M_2$ of size $N\times N$ with a probability distribution given by
the formula 
\bea
{\rm d}\mu(M_1,M_2) = \frac 1{\mathcal Z_N} {\rm d}M_1{\rm d}M_2
\exp\le[-\frac 1 \hbar\tr  \le(V_1(M_1)+V_2(M_2)-M_1M_2\ri)\ri]\
,\nonumber \\
V_1(x) = \sum_{K=1}^\infty \frac {u_K}K x^K\ ;\qquad 
V_2(y) = \sum_{J=1}^{\infty}\frac {v_J}J y^J\ ,
\label{measure}
\eea
where $V_i$ are formal power series but soon will be restricted to
polynomials for simplicity. 
The partition function $\mathcal Z_N$ is known to be a $\tau$-function
for the KP hierarchy in each set of deformation parameters
(coefficients of $V_1$ or $V_2$) and to provide solutions of the
two--Toda hierarchy \cite{UT, AvM1, AvM2}.
This model has been previously investigated in the series of paper
\cite{BHE1, BHE2, BHE3} where a duality of spectral curves and
differential systems for the relevant biorthogonal polynomials has
been unveiled and analyzed in the case of polynomial potentials. In
\cite{BE} the mixed correlation functions of the model (traces of
powers of the two non-commuting matrices) have been reduced to a
formal Fredholm-like determinant without any assumption on the nature
of the potentials and using the recursion coefficients for the
biorthogonal polynomials.
We briefly recall that the biorthogonal polynomials are two sequences
of monic polynomials  (\cite{BHE1} and references therein)
\be
\pi_n(x) = x^n + \cdots , \qquad \sigma_n(y)=y^n + \cdots, \qquad n=0,1,\dots
\ee
that are ``orthogonal'' (better say ``dual'') w.r.t. to the coupled
measure  on the product space
\be
\int_\R\!\!\int_\R\!\!\! {\rm d}x\, {\rm d}y \,\, \pi_n(x)\sigma_m(y) {\rm
e}^{- \frac 1\hbar(V_1(x)+ V_2(y) -xy)} = h_n\delta_{mn} ,\qquad h_n\neq 0\ \forall n\in\N\label{norms}
\ee
where $V_1(x)$ and $V_2(y)$ are  the functions (called {\em potentials})
appearing in the two-matrix model measure (\ref{measure}).
It is convenient to introduce the associated quasipolynomials
 defined by the formulas
\bea
&&\psi_n(x):= \frac 1{\sqrt{h_{n-1}}}\pi_{n-1}(x){\rm e}^{-\frac 1
  \hbar V_1(x)}\\
&&\phi_n(y):= \frac 1{\sqrt{h_{n-1}}}\sigma_{n-1}(y){\rm e}^{-\frac 1
  \hbar V_2(y)}\ .\label{quasidifferentials}
\eea
In terms of these two sequences of quasipolynomials the multiplications
by $x$ and $y$ respectively are represented by semiinfinite square
matrices $Q = [Q_{ij}]_{i,j\in \mathbb N^*}$ and  $P = [P_{ij}]_{i,j\in
  \mathbb N^*}$ according to the formulae
\bea
&&x\psi_n(x) = \sum_m Q_{n,m}\psi_m(x)\ ;\ \ \ y\phi_n(y) = \sum_m
P_{m,n}\phi_m(y) \nonumber \\
&&Q_{n,m}=0=P_{m,n}, \ {\rm if}\ n>m+1.\label{mulxy}
\eea
The matrices $P$ and $Q$ have a rich structure and satisfy the
``string equation'' 
\be
[P,Q]=\hbar \1\ \label{string}\ .
\ee
 We refer for further details to \cite{Berto,BHE1,BHE2,BHE3} where
these models are studied especially in the case of polynomial potentials. We 
also point out that the model can easily be generalized to accommodate
contours of integration other than the real axes \cite{Berto,BHE1}.

The partition function is believed to have a large $N$ expansion 
according to the formula
\be
-\frac 1 {N^2} \ln\mathcal Z_N = \mathcal F =  \mathcal F^{(0)} + \frac 1
{N^2} \mathcal F^{(1)} + \cdots\ .
\ee
This expansion in powers of $N^{-2}$ has been repeatedly advocated for
the $2$-matrix model on the basis of physical arguments \cite{ZJDFG, eynard2}
and has been rigorously proven in the one-matrix model \cite{ercol}.
In the two-matrix model this expansion is believed to generate
2-dimensional statistical models of surfaces triangulated with
ribbon-graphs \cite{ZJDFG, Matrixsurf, Kazakov}, where the powers of
$N^{-1}$ are the Euler characteristics of the surfaces being
tessellated.
From this point of view the term $\mathcal F^{(0)}$ corresponds to a genus
$0$ tessellation and the next to a genus one tessellation.\par
The object of this paper is the leading term of the free energy,
$\mathcal F^{(0)}$. It is the generating function of the expectations of
the powers of the two matrices in the model
\be
\langle {M_1}^K\rangle = K\pa_{u_K} \mathcal F^{(0)} + \mathcal
O(N^{-2})\ ,\qquad  \langle {M_2}^K\rangle = J\pa_{v_J} \mathcal F^{(0)} + \mathcal
O(N^{-2})\ .
\ee
Integral formulas for these partial derivatives at the leading order
are known and involve integrals over a certain spectral curve (see
below \cite{curve}), however a closed formula for the function $\mathcal F^{(0)}$
itself was so far missing; this paper fills the gap
(Thm. \ref{free}). 
Remarkably, an algorithm for the computation of the subleading terms
is known and also a closed expression of the genus $1$
correction $\mathcal F^{(1)}$  \cite{eynard2}, and therefore this
paper precedes logically and complements \cite{eynard2}.
The paper is organized as follows: in section \ref{abelian} we recall the main
formulas known in the literature and set up the notation, linking our
result with the relevant other approaches 
\cite{tt1, tt2, tt3, Teo, WZ, MWZ1, KKWZ,  Zab, Tak}.
The core of the paper is section \ref{freesec} where the formula for
the leading term of the free energy (dToda tau function) is presented
and proved (Thm. \ref{free}). In section \ref{cansec} we apply our result to the
generating function of the canonical change of coordinates represented
by the Lax operators of the dToda hierarchy.\par
Finally in section \ref{genuss} we extend the result obtained in
section \ref{freesec}  and find an integral expression for
the free energy of the two  matrix model in the case the spectral
curve is of arbitrary genus (Thm. \ref{freegenus}).

\section{Planar limit (dToda hierarchy)} 
\label{abelian}
In this section we investigate the planar free energy ($\mathcal
F^{(0)}$ in the notation of the introduction) and we will make soon
the common ``one-cut'' assumption (to be lifted in section
\ref{genuss}) {} which amounts to saying that the
multiplication operators tend to meromorphic functions over a spectral
curve of genus zero. Indeed in this limit the two
multiplication operators for the wave-vectors defined by the
biorthogonal quasipolynomials become commuting functions in the shift
operator  here replaced by the variable $\lambda$ (\cite{BHE3} and
references therein) and they corresponds to the Lax operators of the dToda hierarchy:
\bea
Q(\lambda) = \gamma\lambda + \sum_{k=0}^\infty\alpha_j \lambda^{-j}\\
P(\lambda) = \frac \gamma\lambda + \sum_{j=0}^\infty \beta_j \lambda^j\ .
\eea
The original noncommutativity of the operators now translates to the
following Poisson-bracket which is nothing but the
dispersionless form of the string equation (\ref{string}) \cite{tt1,tt2,tt3}
\be
\le\{P,Q\ri\} := \lambda\frac {\pa P}{\pa\lambda}  \frac{\pa Q}{\pa t} - \lambda\frac
      {\pa Q}{\pa\lambda}  \frac{\pa P}{\pa t}=1\ ,
\label{dToda}
\ee
where $t= \hbar N$ in the large limit.
The parameter $t$ is the (scaled) number of eigenvalues of the matrix
model and enters the relations
\bea
\frac 1 {2i\pi} \oint P {\rm d}Q = t \ ;\ \ 
\frac 1 {2i\pi} \oint Q {\rm d}P = t \ ;\label{normalize}
\eea
The contour chosen is a contour on the physical sheet of the $Q$ ($P$ respectively) plane
around infinity (i.e. around $\lambda =\infty$, $\lambda=0$
respectively).\par
The {\em deformation equations} describe the infinitesimal variations
of the operators  $P,Q$ under variations of the parameters of the
potentials; they are known in the finite $N$ regime as well
(\cite{BHE1, BHE3} and references therein) whereas in the
dispersionless limit are given by the evolution equations \cite{tt1,
  tt2, tt3, Teo}
\bea
(\pa_{u_K} Q)_\lambda = \le\{ Q, (Q^K)_{+0}\ri\}\ ;\qquad
(\pa_{v_J} Q)_\lambda = \le\{Q, (P^J)_{-0}\ri\}\\
(\pa_{u_K} P)_\lambda = \le\{ P, (Q^K)_{+0}\ri\}\ ;\qquad
(\pa_{v_J} P)_\lambda = \le\{P, (P^J)_{-0}\ri\}
\eea
where the subscript $_{0\pm}$ denotes the positive (negative) part of
the Laurent polynomial plus half the part constant in $\lambda$, viz,
e.g.
\be
(Q^K)_{+0}(\lambda) = (Q^K)_{+}(\lambda) + \frac1 2 (Q^K)_0\ .
\ee

If the potentials $V_i$  are polynomials of degrees $d_i+1$, $i=1,2$
then both $P$ and $Q$ are finite Laurent polynomials 
\bea
Q(\lambda) = \gamma\lambda + \sum_{k=0}^{d_2}\alpha_j \lambda^{-j}\\
P(\lambda) = \frac \gamma\lambda + \sum_{j=0}^{d_1} \beta_j \lambda^j\ .
\eea
In what follows we restrict to this case so as to avoid complication
of convergence; however one may replace the contour integrals that
will follow with formal residues of formal Laurent series and carry
out the same computations.\\
Another reason why we prefer the truncated setting is that then the
two functions $P,Q$ define a (singular) spectral curve of genus $g=0$
which is given by the polynomial locus (resultant-like) obtained from
the determinant of the following Sylvester matrix (\cite{BHE2}, thanks
to a remark by J. Hurtubise)
\be
 0 = E(P,Q) =\frac 1{\gamma^{d_1+d_2}} \det\le[\begin{array}{ccccccccc}
\g&\b_0\!-\!P & \b_1&\cdots &\cdots &\b_{d_1}& 0 &0&0\\
0&\g&\b_0\!-\!P & \b_1&\cdots &\cdots &\b_{d_1}& 0 &0\\
0&0&\g&\b_0\!-\!P & \b_1&\cdots &\cdots &\b_{d_1}& 0\\
0&0&0&\g&\b_0\!-\!P & \b_1&\cdots &\cdots &\b_{d_1}\\
\hline
\a_{d_2}&\cdots&\a_1&\a_0\!-\!Q&\g&0&0&0&0\\
0&\a_{d_2}&\cdots&\a_1&\a_0\!-\!Q&\g&0&0&0\\
0&0&\a_{d_2}&\cdots&\a_1&\a_0\!-\!Q&\g&0&0\\
0&0&0&\a_{d_2}&\cdots&\a_1&\a_0\!-\!Q&\g&0\\
0&0&0&0&\a_{d_2}&\cdots&\a_1&\a_0\!-\!Q&\g\\
\end{array}
\ri]\ .
\ee
This spectral curve is precisely the (limit of the) spectral curve of
the four finite-dimensional folded differential systems for the
quasipolynomials \cite{BHE1,BHE3}. If we worked with formal power
series it is not known whether a spectral curve in this sense can be defined.\par

Taking this viewpoint  $\lambda\in \C \P^1$ is
the uniformizing parameter of the spectral curve $E(P,Q)$. The two
potentials are related to the parameters $\g, \a_j,\b_j$ by the relations 
\bea
P = V'_1(Q) -\frac t Q  + \mathcal O(Q^{-2})\ ,\ \ \hbox{near $\infty_Q$}\\
Q = V'_2(P) -\frac t P + \mathcal O(P^{-2})\ ,\ \ \hbox{near $\infty_P$}
\eea
where the point $\infty_Q$ ($\infty_P$) is  the point on the
spectral curve where $Q$ ($P$ respectively) has a simple pole; in the
uniformization provided by the coordinate $\lambda$ it corresponds to
$\lambda=\infty$ ($\lambda =0$ respectively).
By expanding both sides in powers of $\lambda$ and matching the
coefficients one can realize that the coefficients of the two
potentials are rational functions of the parameters $\gamma, \alpha_j, \beta_j$.
More explicitly we have 
\bea
V_1(q) = \sum_{K=1}^{d_1+1} \frac {u_K}{K} q^K\ ;\qquad 
V_2(p) = \sum_{J=1}^{d_2+1} \frac {v_J}J p^J\ ,\\
u_K= -\frac 1{2i\pi}\oint \frac P{Q^K}{\rm d}Q\ ,\qquad  v_J= -\frac
1{2i\pi}\oint \frac Q{P^J}{\rm d}P\ .\label{moments}
\eea
The leading term of the free energy  of the model is then  defined by
the differential equations 
\bea
&&\frac {\pa \mathcal F}{\pa u_K} = U_K:=  \frac 1 {K2i\pi}\oint P\,Q^K {\rm
  d}Q = \frac 1 K \res{\infty_Q} PQ^K{\rm d}Q\nonumber \\
&&\frac {\pa \mathcal F}{\pa v_J} = V_J:= \frac 1 {J2i\pi}\oint Q\,P^J
    {\rm d}P = \frac  1J \res{\infty_P} QP^J{\rm d}P\ .\label{partials}
\eea

These equations are precisely the same that define the $\tau$-function
of the dToda hierarchy. In the relevant literature \cite{tt1, tt2,
  tt3, Teo} the functions $P,Q$ are the Lax operators denoted by $L, \tilde L$ or $\mathcal
L, \tilde{\mathcal L}$ and the normalization is slightly different.

We should also remark the following link to the works \cite{MWZ1,
 KKWZ, WZ, Zab, Tak}
 in that if we take $V_1= V = \overline{V_2}$ we then have $\gamma\in
\R$, $\alpha_j=\overline \beta_j$. The function $Q(\lambda)$ is then
the uniformizing map of a Jordan curve $\Gamma$ in the $Q$-plane (at
 least for suitable ranges of the parameters) which is defined by either of the following relations
\be
P(1/\lambda) = Q(\overline\lambda)\ ; |\lambda| = 1\ .
\ee
 In the setting of \cite{WZ, KKWZ, MWZ1}  the function $Q$ is denoted
 by $z$ (and $\lambda$ by $w$) so that  then $P$ is nothing but the Schwartz function of the curve
 $\Gamma$, defined  by 
\be
\overline z = S(z)\ ,  \ \ z\in \Gamma\ .
\ee 
The coefficients of the potential $V(x) =
\sum_{k}\frac {t^k}kx^k$ are the so-called ``exterior harmonic
moments'' of the region $\mathcal D$ enclosed by the curve $\Gamma$
\bea
&& t_K =
\frac 1 {2i\pi}\oint_{\Gamma} \overline z z^{-K}{\rm d}z\\
&& t=t_0 = \frac 1{2i\pi} \int\!\!\!\!\!\int_{\mathcal D} {\rm
  d}z\wedge{\rm d}\overline z =  \frac 1{2i\pi}\oint_{\Gamma}
\overline z {\rm d}z= \frac{Area(\mathcal D)}\pi\ .
\eea
By writing $\overline z=S(z(w))$ these integrals become residues in
the $w$-plane. 
In our general situation the representation of the
coefficients $u_K, v_J$ (\ref{moments}) cannot be translated to a surface integral,
hence the need for a separate analysis.
For conformal maps (i.e. Jordan curves)  the $\tau$-function has been defined in \cite{Tak}
and given an appealing interpretation as (exponential of the Legendre transform of) the
electrostatic potential of a uniform $2$-dimensional distribution of charge in
$\mathcal D$ \cite{KKWZ} 
\be
\ln(\tau_\Gamma) = -\frac 1 \pi \int\!\!\!\int_{\mathcal D} \ln\le|\frac 1 z
- \frac 1 {z'} \ri|{\rm d}^2 z{\rm d}^2 z'\ .
\ee
It can be rewritten as a (formal) series in the exterior and interior
 moments as 
\be
2\ln(\tau_\Gamma) = -\frac 1 {4\pi} \int\!\!\!\!\int_{\mathcal D} {\rm
 d}^2 z |z|^2 + t_0w_0 +\sum_{K>0}(t_K w_K+ \overline {t}_K \overline{w}_K)
\ee
where the interior moments are defined by (the normalization here
 differs slightly from \cite{WZ})
\bea
w_K =\frac 1 {\pi K}  \int\!\!\!\!\int _{\mathcal D}z^K {\rm d}^2z\ ,\
 \ K>0\\
w_0 = \frac 1{\pi}  \int\!\!\!\!\int _{\mathcal D} \ln|z|^2{\rm d}^2z\ .
\eea
Unfortunately this formula is not exportable to our more general
setting 
(in particular the logarithmic moment above) and to the general setting of the dispersionless Toda hierarchy;
our principal objective is to fill this gap.
\subsection{Free energy of the $2$-matrix model in the planar limit}
\label{freesec}
Let us focus on the function $P$ as a (multivalued) function of $Q$
(similar argument can be reversed for $Q$ as a function of $P$);
on the physical sheet $P(Q)$ will have in general some branch-cuts with
square-root singularities at the endpoint of each cut. 
These cuts are bounded in the physical sheet because $P(Q)$ is
analytic in a neighborhood of $\infty_Q$. Note that 
\bea
V_1(q) =\frac 1 {2i\pi}\oint \ln\le(1-\frac q Q\ri)P{\rm d}Q\\
V_2(p) = \frac 1 {2i\pi}\oint \ln\le(1-\frac p P\ri)Q{\rm d}P\ ,
\eea
as one can immediately realize by expanding in powers of $q,p$. 
The integrals are well defined provided that $q$ ($p$
respectively) are kept inside the contour of integration. In the
following we will
develop all the necessary arguments only for $V_1(q)$ and related
objects, where the reader can obtain the relevant proofs for $V_2(p)$
by interchanging the r\^ole of $P$ and $Q$.\par

Let us now introduce the {\em exterior potentials}\footnote{
It has been pointed out to me by B. Eynard that they
 are sometimes referred to as {\em effective} potentials because
they can be thought of as  the effective potential felt by one eigenvalue
in the field of the others.}
\be
\Phi_1 (q_{out}) = - \frac 1 {2i\pi} \oint  \ln\le(1-\frac Q {q_{out}}\ri)P{\rm d}Q \label{outer}
\ee
In this formula the contour of integration is such as to leave the
point $q_{out}$ in the outside region (whence the subscript). In
general we will distinguish the choice of the contours by a subscript
$q_{out}$ or $q_{in}$ in what follows.\\
The coefficients of $\Phi_1(q_{out})$ in inverse powers of $q_{out}$
are precisely (minus) the $U_K$ defined in (\ref{partials}). 
Note that expanding eq. (\ref{outer}) in inverse powers of $q_{out}$
the first $d_1+1$ coefficients are the $U_K$ coefficients as
defined in (\ref{partials}).\par 
The first objective is to analytically continue $\Phi_1$ to the physical
sheet so as to  obtain a different representation of it.
For this purpose we compute
\bea
 && \Phi_1'(q_{out}) =  \frac 1 {2i\pi} \oint  \frac Q
    {q_{out}(Q-q_{out})} P{\rm d}Q =   \frac {  \oint P{\rm d}Q}{q_{out}2i\pi} + \frac 1 {2i\pi} \oint  \frac 1 {(Q-q_{out})
}P{\rm d}Q = \\
&&= \frac t q  +\frac 1 {2i\pi} \oint  \frac 1 {Q-q_{in}}P{\rm d}Q +
    P(q) =
 \frac t q  -V_1'(q) +
    P(q) \ , 
\eea
where we have dropped the subscript outside of the integral as those terms
are analytic functions in the whole physical sheet. Integrating once
we obtain
\be
\Phi_1(Q) = -V_1(Q)+ t\,\ln (Q)  +\int_{X_q} \!\!\!P{\rm d}Q\ ,
\ee
where $X_q$ is a point defined implicitly by the requirement $\Phi_1 = \mathcal O(Q^{-1})$
near $\infty_Q(\equiv(\lambda=\infty))$.
Note that, in spite of the $\ln(Q)$ term, this is an analytic function
around $\infty_Q$.\\
By similar reasoning we get
\be
\Phi_2(P) = -\frac 1{2i\pi} \oint \ln\le(1-\frac
{\tilde P}{P}\ri)\tilde Q{\rm d}\tilde P = -V_2(P)+t\,\ln(P)+ \int_{X_p}\!\!\!Q{\rm d}P
\ee
We need to introduce two more points
(beside $X_q$ and $X_p$) on the
spectral curve and a lemma: those are the points  $\Lambda_q,\
\Lambda_p$ defined implicitly by the relations 
\bea
 && \int_{\Lambda_q}\!\!\! P{\rm d}Q = V_1(Q(\lambda))_{>0}+ t\,\ln(\lambda) 
+\mathcal O(\lambda^{-1})\ ,\ \
\hbox{near $\infty_Q$} \\
&&\int_{\Lambda_p} \!\!\! Q{\rm d}P = V_2(P(\lambda))_{<0}+  t\,\ln(\lambda)
+\mathcal O(\lambda)\ ,\ \
\hbox{near $\infty_P$}\ .
\eea
By a simple inspection of the $\lambda^0$--coefficient in the  LHS and RHS one finds that
\bea
 && \int_{X_q}\!\!\! P{\rm d}Q =\int_{\Lambda_q}\!\!\!P{\rm d}Q  +
\int^{\Lambda_q}_{X_q}\!\!\! P{\rm d}Q = \int_{\Lambda_q}\!\!\! P{\rm d}Q +
(V_1(Q))_0 - t\, \ln(\gamma)\label{uno}\\
&& \int_{X_p}\!\!\! Q{\rm d}P =\int_{\Lambda_p}\!\!\! Q{\rm d}P  +
\int^{\Lambda_p}_{X_p}\!\!\! Q{\rm d}P = \int_{\Lambda_p}\!\!\! Q{\rm d}P +
(V_2(P))_0 -  t\,\ln(\gamma)\ ,\label{due}
\eea
where the subscript $_0$ denotes the constant part in $\lambda$ (which
can be written as a residue).\par
We now have the 
\bl
\label{lemma1}
The following relation holds
\be
\mu:= Q(X_p)P(X_p)+\int_{X_p}^{X_q}\!\!\!\! P{\rm d}Q 
 =Q(X_q)P(X_q)+\int_{X_q}^{X_p}\!\!\!\! Q{\rm d}P =
\big(PQ-V_1(Q)-V_2(P)\big)_0 +2 t\,\ln(\gamma)\ .
\ee
\el
\br
The quantity $\mu$ will be proved in Corollary \ref{Ft} to be the
derivative of the free energy w.r.t $t$, i.e. the chemical
potential. It therefore  corresponds to the logarithmic moment in the setting of
\cite{WZ, MWZ1, KKWZ, Zab}.
\er
{\bf Proof.} The equivalence of the two integrals is immediate by an
integration by parts. And integration by parts is in fact the key to
prove also the last part:
\bea
&& \int_{\Lambda_q}\!\!\!\!P{\rm d}Q = PQ-P(\Lambda_q)Q(\Lambda_q) -
\int_{\Lambda_q} \!\!\!\!Q{\rm d}P =  PQ- P(\Lambda_q)Q(\Lambda_q) -\int_{\Lambda_q}^{\Lambda_p}
\!\!\!\! Q{\rm
  d}P -\int_{\Lambda_p} \!\!\!\!Q{\rm d}P\ .
\eea
Looking at the constant part in $\lambda$ in both sides of the above equation we conclude that 
\be
(QP)_0= P(\Lambda_q)Q(\Lambda_q) +\int_{\Lambda_q}^{\Lambda_p}\!\!\!\! Q{\rm
  d}P\ .
\ee
Therefore we have
\bea
(QP)_0 \hspace{-18pt} &&= P(\Lambda_q)Q(\Lambda_q) +\int_{\Lambda_q}^{\Lambda_p} \!\!\!\!Q{\rm  d}P =
  P(\Lambda_q)Q(\Lambda_q) +\int_{X_p}^{\Lambda_p}\!\!\!\! Q{\rm d}P+ 
\int_{\Lambda_q}^{X_p}\!\!\!\! Q{\rm  d}P =\\
 &&= 
  P(\Lambda_q)Q(\Lambda_q) +\int_{X_p}^{\Lambda_p} \!\!\!\!Q{\rm d}P+ 
Q(X_p)P(X_p)-Q(\Lambda_q)P(\Lambda_q) -\int_{\Lambda_q}^{X_p}\!\!\!\! P{\rm
  d}Q=\\
&&= \int_{X_p}^{\Lambda_p}\!\!\!\! Q{\rm d}P+ 
Q(X_p)P(X_p) -\int_{X_q}^{X_p}\!\!\!\! P{\rm
  d}Q- \int_{\Lambda_q}^{X_q}\!\!\!\! P{\rm
  d}Q =\\
&& =\big(V_1(Q)+V_2(P)\big)_0-2 t\,\ln(\gamma)+
Q(X_p)P(X_p) +\int_{X_p}^{X_q} \!\!\!\!P{\rm
  d}Q\ .
\eea
This concludes the proof of the Lemma. Q.E.D.\par\vskip 3pt

We are now in a position to formulate the first main result of this paper.

\bt
\label{free}
The Free energy is given by the formula (up to constant)
\bea
2\mathcal F \!\!\!&=&\!\!\! \res{\infty_Q} \le(P\Phi_1{\rm d}Q \ri)\!+
\!\!\res{\infty_P}\le(Q\Phi_2{\rm d}P\ri)\! +\!\!\frac 1 2 
\res{\infty_Q} \le(P^2 Q{\rm d}Q\ri) + t \res{\lambda=\infty} \le(\frac {V_1(Q)+V_2(P)-PQ}{\lambda}{\rm
  d}\lambda\ri) \!+ \!2t^2\ln(\gamma)\label{T0}
\\
&=& \!\!\! \sum_{K}u_KU_K +  \sum_J v_JV_J  +\frac 1 2
\res{\infty_Q} P^2 Q{\rm d}Q + t \res{\lambda=\infty}\le(\frac {V_1(Q)+V_2(P)-PQ}{\lambda}{\rm
  d}\lambda\ri) + 2 t^2 \ln(\gamma)
\ . \label{freeen}
\eea
\et
Before proceeding to the proof a corollary  and some remarks are in
order. First off from the expressions in Thm. \ref{free} we find the
well-known scaling property of the free energy \cite{KKWZ,Zab,Boya}
\bc
\label{freecor}
The free energy defined in Thm. \ref{free} satisfies the scaling equation
\be
4\mathcal F =-t^2 + \bigg( 2t\pa_t + \sum_K (2-K)u_K\pa_{u_K} +
\sum_J(2-J) v_J\pa_{v_J}\bigg)\mathcal F\ .\label{scall}
\ee
(More general scaling equations will be introduced later in Corollary \ref{scalg}).
\ec
{\bf Proof}.
The proof in the context of the normal matrix model can be found in
the references quoted above and, in view of the formal equivalence of
the normal matrix model with the two--matrix model \cite{KKWZ}, the
statement would follow also in our case.\par 
In order to be self contained we rederive this property in the present
context; one way of proving formula (\ref{scall}) is  from the expression (\ref{freeen}) for $\mathcal F$
given in Thm. \ref{free}, by computing the residues involved after
rewriting symmetrically the term
$\res{\infty_Q} P^2Q{\rm d}Q $  as $\frac 12 \le(\res{\infty_Q} P^2Q{\rm
  d}Q+  \res{\infty_P} Q^2P{\rm d}P\ri)$ and computing it.
A second, possibly  instructive way to derive it also 
directly from eq. (\ref{T0}) is by  using the scaling properties of the various quantities
involved.
 To this purpose we introduce the rescaling according to the formul\ae\ 
\be
Q = \delta \widetilde Q\ ,\qquad P = \delta \widetilde P\ . \label{conf1}
\ee
Under this change of the functions $Q,P$ we have (using formulas
(\ref{normalize}, \ref{partials}))
\be
u_K = \delta^{2-K} \widetilde u_K\ ;\ \ v_J = \delta^{2-J}\widetilde
v_J\ ;\ \ t = \delta^2 \widetilde t\ .\label{conf2}
\ee
Moreover, computing explicitly the residue $\res{\lambda=\infty}
PQ'{\rm d}\lambda$ we find that 
\be
\gamma^2 = -t+\sum_j j\alpha_j\beta_j\ ,\label{conf3}
\ee
from which one immediately obtains  that  $\gamma = \delta \widetilde
\gamma$.
 The exterior potentials are also conformally invariant, indeed 
\bea
\Phi_1\!\!\! &=&\!\!\! -V_1(Q)+t\ln(Q) +\!\!\int_{X_q}\!\!\!\! P{\rm d}Q = 
-\delta^2\widetilde V_1(\widetilde Q) + \delta^2 \widetilde
t\ln(\widetilde Q) +\!\! \int_{\widetilde{X_q}} \!\!\!\!\widetilde P{\rm
  d}\widetilde Q + \delta^2\widetilde t \ln (\delta)
+\int_{X_q}^{\widetilde {X_q}}\!\!\!\!\widetilde P{\rm d}\widetilde Q
\\
&=&\!\!\!\delta^2 \widetilde{\Phi_1} +  \delta^2\widetilde t \ln (\delta)
+\int_{X_q}^{\widetilde {X_q}}\!\!\!\!\widetilde P{\rm d}\widetilde Q
\eea
The last two terms cancel each other out because both the LHS and RHS
must be $\mathcal O(Q^{-1}) = \mathcal O(\widetilde Q^{-1})$.
Repeating the argument for $\Phi_2$ we finally get 
\be
\Phi_1 = \delta^2 \widetilde{\Phi_1}\ ;\ \ \Phi_2 = \delta^2
\widetilde {\Phi_2}\ .\label{conf4}
\ee
Plugging the relations (\ref{conf1}, \ref{conf2}, \ref{conf3},
\ref{conf4})  into the expression (\ref{freeen})  for the free energy we obtain
\be
\mathcal F = \delta ^4\widetilde{ \mathcal F}  + \delta^4 \widetilde
t^2 \ln \delta = \delta ^4\widetilde{ \mathcal F} + t^2\ln(\delta)
\ee
Applying the operator $\delta\pa_\delta|_{\delta=1}$ to both sides we
obtain
\be
0 = 4\mathcal F - \le[\sum_K(2-K)u_K\pa_{u_K} + \sum_J (2-J)v_J\pa_{v_J} + 2
  t\pa_t\ri] \mathcal F + t^2\ ,
\ee
whence the statement of the corollary. Q.E.D.\par\vskip 4pt
We also add a few remarks before moving on with the proof of
Thm. \ref{free}.
\br
Formulas (\ref{T0}, \ref{freeen}) are symmetric in the r\^oles of $P$ and $Q$: the only non-immediately
symmetric term is $\res{\infty_Q} P^2Q{\rm d}Q $ but an integration by parts restores the symmetry.
\er
\br
The formula is derived for polynomial potentials but it could
possibly be extended to convergent or formal series.
\er
\br
The genus $1$ correction to the above formula has been computed in
\cite{eynard2} and is given by 
\bea
\mathcal F^{(1)} \hspace{-18pt}&& = -\frac 1 {24}\ln (\gamma^4 D)\\
&&D = \frac 1 {\gamma^{d_1+d_2+2}} \det\le[
\begin{array}{ccccccccc}
-\g&0 & \b_1&\cdots &\cdots &d_1\b_{d_1}& 0 &0&0\\
0&-\g&0 & \b_1&\cdots &\cdots &d_1\b_{d_1}& 0 &0\\
0&0&-\g&0 & \b_1&\cdots &\cdots &d_1\b_{d_1}& 0\\
0&0&0&-\g&0 & \b_1&\cdots &\cdots &d_1\b_{d_1}\\
\hline
d_2\a_{d_2}&\cdots&\a_1&0&-\g&0&0&0&0\\
0&d_2\a_{d_2}&\cdots&\a_1&0&-\g&0&0&0\\
0&0&d_2\a_{d_2}&\cdots&\a_1&0&-\g&0&0\\
0&0&0&d_2\a_{d_2}&\cdots&\a_1&0&-\g&0\\
0&0&0&0&d_2\a_{d_2}&\cdots&\a_1&0&-\g\\
\end{array}
\ri] 
\eea
\er
{\bf Proof of  Theorem \ref{free}.}
The equivalence of lines (\ref{T0}) and (\ref{freeen}) follows from the definition of the
exterior potentials and Lemma \ref{lemma1}.\\
Let us consider the derivative $\pa_K:= \frac \pa{\pa u_K}$ done at
argument ($P$ or $Q$) fixed, where the value being kept fixed under
differentiation is denoted by the corresponding subscript
\bea 
 (\pa_K\Phi_1)_Q\!\!\! &=&\!\!\! -\frac {Q^K}K+
\int_{X_q} \!\!\!(\pa_K P)_Q{\rm d}Q -P(X_q)\pa_K(Q(X_q)) \\
 (\pa_K\Phi_2)_P\!\!\! &=&\!\!\! \int_{X_p} \!\!\!(\pa_K Q)_P{\rm d}P-Q(X_p)\pa_K (P(X_p))
\eea
Let us set 
\be
4i\pi \mathcal F_0:=  \oint_{\infty_Q}\!\!\!\!\! P\Phi_1(Q){\rm d}Q +
\oint_{\infty_P}\!\!\!\!\! Q\Phi_2(P){\rm d}P +\frac  1 2
\oint_{\infty_Q}\!\!\!\!\!P^2 Q{\rm d}Q \ ,\label{F0}
\ee
and study the variation of this functional.
First off we have the formulas
\bea
P \!\!\!&=&\!\!\! V_1'(Q) -\frac t Q +\sum_{K=1}^\infty K U_K Q^{-K-1}\\
\Phi_1\!\!\! &=&\!\!\! -\sum_{K=1}^\infty U_K Q^{-K}\\
Q\!\!\! &=& \!\!\!V_2'(P)- \frac t P +\sum_{J=1}^\infty J V_J P^{-J-1} \\
\Phi_2\!\!\! &=&\!\!\! -\sum_{J=1}^\infty V_J P^{-J}\ ,
\eea
where the $U_K, V_J$'s have been defined in (\ref{partials}).
We also need the variations of the functions $P,Q$ given here below
\bea
&& (\pa_K P)_Q = Q^{K-1} + \mathcal O(Q^{-2})\ ;\qquad  (\pa_KQ)_P = \mathcal O(P^{-2})\ .
\eea
With these formul\ae\ we can now compute the variation of $\mathcal
F_0$ (the subscript on the loop integrals that follow specify the
points around which we circulate):
\bea
&& 4i\pi\pa_K \mathcal F_0 = \overbrace{\oint_{\infty_Q}\!\!\!\! (\pa_K P)_Q\,\Phi_1{\rm
    d}Q}^{=2i\pi U_K}
\hspace{-1pt}+
\hspace{-1pt}
\overbrace{\oint_{\infty_Q}\!\!\!\!\!
P\,(\pa_K\Phi_1)_Q{\rm d}Q}^{ =\star} 
\hspace{-0pt} + 
\hspace{-4pt}
\overbrace{\oint_{\infty_P}\!\!\!\!\! (\pa_K Q)_P\,\Phi_2{\rm d}P}^{=0} +\oint_{\infty_P}\!\!\!\!\!
Q\,(\pa_K\Phi_2)_P{\rm d}P + \oint_{\infty_Q}\!\!\!\!\! P(\pa_K P)_Q Q{\rm d}Q =
\\
&&=4i\pi U_k +  \oint_{\infty_Q}\!\!\!\!\!  P \le[\int_{X_q}\!\!\! (\pa_k P)_Q{\rm d}Q\ri] {\rm d}Q + 
\oint_{\infty_P}\!\!\!\!\!  Q \le[\int_{X_p} \!\!\!(\pa_k Q)_P{\rm d}P\ri] {\rm d}P  +  \oint_{\infty_Q}\!\!\!\!\! P(\pa_K P
)_Q Q{\rm d}Q +\\
&&-2i\pi t\le[P(X_q)\pa_K(Q(X_q)) +Q(X_p)\pa_K (P(X_p))\ri] = \\
&&= 4i\pi U_k +  \oint_{\infty_Q}\!\!\!\!\! P \le[\int_{X_q}\!\!\! (\pa_k P)_Q{\rm d}Q\ri] {\rm d}Q  + 
\oint_{\infty_P}\!\!\!\!\!  Q \le[\int_{X_q} \!\!\!(\pa_k Q)_P{\rm d}P\ri] {\rm d}P
+\\
&&  -
\oint_{\infty_Q}\!\!\!\!\! P \le[\int_{X_q}\!\!\! (\pa_k P)_Q{\rm d}Q\ri] {\rm d}Q  - 
\oint_{\infty_Q}\!\!\!\!\! Q \le[\int_{X_q}\!\!\! (\pa_k P)_Q{\rm d}Q\ri] {\rm d}P\label{rrr}+\\
&&
-2i\pi t\le[P(X_q)\pa_K(Q(X_q)) +Q(X_p)\pa_K (P(X_p)) -
  \int_{X_p}^{X_q}(\pa_KQ)_P{\rm d}P\ri]
\eea
where
\be
\star = 2i\pi U_k + \oint_{\infty_Q}\!\!\!\!\!
  P\le[\int_{X_q}\!\!\!\! (\pa_k P)_Q{\rm d}Q\ri] {\rm d}Q -2i\pi t P(X_q)\pa_K(Q(X_q))\ .
\ee
We now use the ``thermodynamic identity''
\be
(\pa_K P)_Q {\rm d}Q = -(\pa_K Q)_P{\rm d}P\ ,\label{thermos}
\ee
so that the last term on line (\ref{rrr}) reads (notice the double change of sign due to
the definition of the circle around $\infty_Q$, which  is (homologically)
minus the circle around $\infty_P$)
\bea
&& \oint_{\infty_Q}\!\!\!\!\! Q \le[\int_{X_q}\!\!\! (\pa_k P)_Q{\rm d}Q\ri] {\rm d}P  = 
\oint_{\infty_P}\!\!\!\!\! Q \le[\int_{X_q}\!\!\! (\pa_k Q)_P{\rm d}P\ri] {\rm d}P   \label{genus}
\eea
Plugging into the variation of $\mathcal F_0$ we get
\be
4i\pi \pa_K\mathcal F_0 =
4i\pi U_K -2i\pi \,t\le[P(X_q)\pa_K(Q(X_q)) +Q(X_p)\pa_K (P(X_p)) -
  \int_{X_p}^{X_q}(\pa_KQ)_P{\rm d}P\ri]
\label{errr}
\ee
Finally we claim that the term in the square brackets in (\ref{errr})
is 
\bea
P(X_q)\pa(Q(X_q)) +Q(X_p)\pa (P(X_p)) -
  \int_{X_p}^{X_q}(\pa Q)_P{\rm d}P =
  \pa\le[Q(X_p)P(X_p)+\int_{X_p}^{X_q} P{\rm d}Q \ri]=\\
=
\pa\le[\bigg(PQ-V_1(Q)-V_2(P)\bigg)_0 +2t\,\ln(\gamma)\ri]\ ,
\eea
where $\pa$ denotes any vector field in the space of parameters
$u_K,v_J$ (or even $t$).
Indeed
\bea
&&\pa\le[Q(X_p)P(X_p)+\int_{X_p}^{X_q}\!\!\!\! P{\rm d}Q \ri] =
\pa\le[Q(X_p)P(X_p)+\int_{X_p}^{X_q}\!\!\!\! P(\lambda)Q'(\lambda){\rm
    d}\lambda \ri] = \\
&&= \pa(Q(X_p))P(X_p)+Q(X_p)\pa(P(X_p)) + \\
&& \hspace {0.4cm}+\int_{X_p}^{X_q}\!\!\!\! \big[(\pa
P)_\lambda(\lambda)Q'(\lambda) + P(\lambda)(\pa Q')_\lambda(\lambda)\big]{\rm
    d}\lambda - P(X_p)Q'(X_p)\pa X_p + P(X_q)Q'(X_q)\pa X_q =
\eea
\bea
&&=
\pa(Q(X_p))P(X_p)+Q(X_p)\pa(P(X_p)) + \\
&& \hspace {0.4cm}+\int_{X_p}^{X_q}\!\!\!\!\big[ (\pa
P)_\lambda(\lambda)Q'(\lambda) - P'(\lambda)(\pa Q)_\lambda(\lambda)
\big]{\rm
    d}\lambda +P(X_q)\pa(Q)_\lambda(X_q) -
P(X_p)\pa(Q)_\lambda(X_p)+\\
&&\hspace {1cm}- P(X_p)Q'(X_p)\pa X_p +P(X_q)Q'(X_q)\pa X_q\ .\label{end1}
\eea
In order to proceed we note that if $X$ is a point depending on the parameters, we have 
\be
\pa(Q(X)) = (\pa Q)_\lambda(X)+Q'(X)\pa X
\ee
Using this formula we get
\bea
\hbox{(\ref{end1})}\hspace{-18pt}&& = Q(X_p)\pa(P(X_p)) + \pa(Q(X_p)) P(X_p) + \int_{X_p}^{X_q}\!\!\!\!\big[ (\pa
P)_\lambda(\lambda)Q'(\lambda) - P'(\lambda)(\pa Q)_\lambda(\lambda)
\big]{\rm
    d}\lambda= \\
&&= Q(X_p)\pa(P(X_p)) + \pa(Q(X_p)) P(X_p) + \int_{X_p}^{X_q}\!\!\!\! (\pa P)_Q
{\rm d}Q\ ,
\eea
which is the term in square brackets in (\ref{errr}).
Using then Lemma \ref{lemma1} we have proven that 
\be
2\pa_K \mathcal F_0 = 2U_K -t  \pa_K\le\{\le[-V_1(Q)-V_2(P)+QP\ri]_0 +
2t\,\ln (\gamma)\ri\} =  2U_K -t  \pa_K\mu\ .
\ee
The second term is exactly the opposite of the variation of the last
two terms in formula (\ref{freeen}), and
this concludes the proof (the derivatives w.r.t. $v_J$ are treated by
interchanging the roles of $P$ and $Q$). Q.E.D. \par\vskip 3pt
The derivative w.r.t the number operator $t$ has to be treated in a
separate way.
\bc\label{Ft}
The derivative of the free energy w.r.t $t$ is given by the formula
\be
\pa_t\mathcal F = Q(X_p)P(X_p)+ \int_{X_p}^{X_q}\!\!\!\!  Q
{\rm d}P = \mu\ .
\ee
\ec
{\bf Proof.} We start with the following observation

\be
(\pa_tP)_Q{\rm d}Q = -(\pa_tQ)_P{\rm d}P = -\frac {{\rm
    d}\lambda}\lambda\ .\label{thirdkind}
\ee
In other words they are  differential of the third kind with poles
at $\infty_Q$ and $\infty_P$ and opposite residues, to wit (using the
thermodynamical identity (\ref{thermos}))
\bea
-\frac 1 P{\rm d}P + \mathcal O(P^{-2}) = (\pa_tQ)_P{\rm d}P  = -
(\pa_tP)_Q{\rm d}Q = \frac 1 Q{\rm d}Q + \mathcal O(Q^{-2})
\eea
and  the differentials have no other singularities.
We then proceed as  in the proof of Thm. \ref{free} 
\bea
 (\pa_t\Phi_1)_Q\!\!\! &=&\!\!\! \ln(Q)+
\int_{X_q} \!\!\!(\pa_t P)_Q{\rm d}Q -P(X_q)\pa_t(Q(X_q)) \\
 (\pa_t\Phi_2)_P\!\!\! &=&\!\!\! \ln(P)+ \int_{X_p} \!\!\!(\pa_t
Q)_P{\rm d}P-Q(X_p)\pa_t (P(X_p))=\\
 &=& \!\!\! \ln(P)+ \int_{X_q} \!\!\!(\pa_t
Q)_P{\rm d}P+ \int_{X_p}^{X_q} \!\!\!(\pa_t
Q)_P{\rm d}P -Q(X_p)\pa_t (P(X_p)) = \\
&=&\!\!\!   \ln(P)+ \int_{X_q} \!\!\!(\pa_t
Q)_P{\rm d}P- \int_{X_p}^{X_q} \!\!\!(\pa_t
P)_Q{\rm d}Q -Q(X_p)\pa_t (P(X_p))\\
(\pa_tP)_Q\!\!\!  &=&\!\!\!  -\frac 1 Q + \mathcal O(Q^{-2})\\
(\pa_tQ)_P\!\!\!  &=&\!\!\!  -\frac 1 P + \mathcal O(P^{-2})\ .
\eea
We then find that 
\bea
4i\pi\pa_t\mathcal F_0\!\!\!\! &=&\!\!\!\! \oint_{\infty_Q}\!\!\!\!\! P\le[ \ln(Q)+
\int_{X_q} \!\!\!\overbrace{(\pa_t P)_Q{\rm d}Q}^{= - {\rm
 d}\lambda/\lambda}\ri] {\rm d}Q + \oint_{\infty_P}\!\!\!\!\! Q\le[ \ln(P)+ \int_{X_q}
 \!\!\! \overbrace{(\pa_t
Q)_P{\rm d}P}^{={\rm d}\lambda/\lambda}\ri] {\rm d}P + \nonumber\\
&& - 2i\pi\,t\pa_t \le[ Q(X_p)P(X_p)+ \int_{X_p}^{X_q}\!\!\!\!  Q
{\rm d}P\ri]+\oint_{\infty_Q}\!\!\!\!\! PQ\overbrace{(\pa_tP)_Q{\rm d}Q}^{={\rm d}\lambda/\lambda}=\\
&=&\!\!\!  - 2i\pi\,t\pa_t \le[ Q(X_p)P(X_p)+ \int_{X_p}^{X_q}\!\!\!\!  Q
{\rm d}P\ri] -2i\pi(PQ)_0+\oint_{\infty_Q}\!\!\!\!\! P\le[ \ln(Q) -\ln(\lambda)\ri] {\rm d}Q+
 \oint_{\infty_P}\!\!\!\!\! Q\le[\ln(P)+\ln(\lambda)\ri]{\rm d}P \label{end2}
\eea
We now claim that 
\be
\oint_{\infty_Q}\!\!\!\!\!\! P\le[ \ln(Q) -\ln(\lambda)\ri]  {\rm d}Q =
 V_1(Q)_0-t\ln(\gamma)\ ,
\ee
and the symmetric formula for the other term. Indeed 
\be
\ln\le(\frac Q \lambda\ri) = \ln\gamma + \mathcal O(Q^{-1})\ ,\ \
 \hbox{near $\infty_Q$}
\ee and thus
\bea
\oint_{\infty_Q}\!\!\!\!\!\! P\le[ \ln(Q) -\ln(\lambda)\ri]  {\rm d}Q\!\!\!\! &=&\!\!\!\!
\oint_{\infty_Q}\!\!\!\! \le(V_1'(Q)-\frac t Q + \mathcal O(Q^{-2})\ri)\le[ \ln(Q)
 -\ln(\lambda)\ri]  {\rm d}Q=\\
&=&\!\!\!\! \oint_{\infty_Q}\!\!\!\!\!\! \le(V_1'(Q) + \mathcal O(Q^{-2})\ri)\le[ \ln(Q)
 -\ln(\lambda)\ri]  {\rm d}Q +2i\pi\, t\ln(\gamma)=\\
&=&\!\!\!\!-\oint_{\infty_Q}\!\!\!\!\!\! \le(V_1(Q) + \mathcal O(Q^{-1})\ri)\le[\frac {{\rm d}Q}Q-\frac
 {{\rm d}\lambda}\lambda\ri]  +2i\pi\, t\ln(\gamma)=\\
&=&\!\!\!\!-2i\pi\,(V_1(Q))_0+ 2i\pi\,t\ln(\gamma)\ .
\eea
Plugging this into (\ref{end2}) we obtain
\be
2 \pa_t\mathcal F_0 = -t\pa_t \le[ Q(X_p)P(X_p)+
 \int_{X_p}^{X_q}\!\!\!\!  Q 
{\rm d}P\ri]+\le[ Q(X_p)P(X_p)+
 \int_{X_p}^{X_q}\!\!\!\!  Q 
{\rm d}P\ri]
\ee
so that 
\be
2 \pa_t\mathcal F  = 2 \le[ Q(X_p)P(X_p)+
 \int_{X_p}^{X_q}\!\!\!\!  Q 
{\rm d}P\ri] \ .
\ee
Q.E.D.\par\vskip 4pt
%
\subsection{Canonical transformations}
\label{cansec}
In the Poisson structure (\ref{dToda}) the coordinates $\ln(\lambda)$
and $t$ are canonically conjugate, as well as the functions
$P,Q$. Therefore it makes sense to find the generating function of
these transformations. This was accomplished in the context of
conformal maps in \cite{WZ} but it is probably a new result in this
context, since now we can express explicitly the generating function
as integrals of the dToda operators $P,Q$.
I recall that we are looking for a function $\Omega(Q,t)$ with the
property that 
\be
d_{Q,t}\Omega = P{\rm d}Q + \ln(\lambda){\rm d}t\ .
\ee
The following proposition gives a representation of such function
\bp
\label{canonic}
The generating function of the canonical change of coordinates
$(\ln(\lambda),t)\mapsto (P,Q)$  is given by  
\bea
\Omega \hspace{-18pt}&& = \int_{X_q} P{\rm d}Q-\frac \mu 2 = \int_{X_q} P{\rm d}Q -\frac 1 2 \le[ Q(X_p)P(X_p)+
 \int_{X_p}^{X_q}\!\!\!\!  P 
{\rm d}Q\ri] = \\
&&= \int_{X_q} P{\rm
  d}Q +\frac 1 2 \big(V_1(Q)+V_2(P)-PQ\big)_0 - t\,\ln(\gamma)
\eea
is the generating function of the canonical
transformation  $(\ln(\lambda), t)\to (P,Q)$, or, 
\be
{\rm d}_{Q,t}\Omega = \pa_Q\Omega_1{\rm d}Q + (\pa_t\Omega)_Q{\rm
  d}t = P{\rm d}Q + \ln(\lambda){\rm d}t\ .
\ee
\ep
\br
The function that we get by interchanging the r\^ole of $P$ and $Q$
would generate the transformation $(-\ln(\lambda),t)\mapsto (Q,P)$ (or
the anti canonical one).
\er
{\bf Proof.}
The first part is obvious
\be
\pa_Q \Omega = P \qquad \hbox{by definition.}
\ee
As for the second we compute
\bea
(\pa_t\Omega)_Q \hspace{-18pt}&&= \pa_t\le( V_1(Q)+t\ln(Q) - \frac 1  2\mu
+ \mathcal O(Q^{-1})\ri)_Q =\\
&&= \ln(Q)-\frac 12  \pa_t\mu + \mathcal O(Q^{-1})=\\
&&=\ln(\lambda)+\ln(\gamma)-\frac 1 2 \pa_t \mu + \mathcal
O(\lambda^{-1})\ .
\eea
It is well known in the theory of the dToda tau-function that 
\be
\pa^2_t\mathcal F = \pa_t\mu = 2 \ln(\gamma)\ ,
\ee 
and hence the constant term in the above expression vanishes.
On the other hand
\bea
(\pa_t\Omega)_Q \hspace{-18pt}&&= \int_{X_q}\!\!\!(\pa_t
P)_Q{\rm d}Q - P(X_q)\pa_t(Q(X_q)) = \int_{X_q}\!\!\!\frac {{\rm
    d}\lambda}\lambda - P(X_q)\pa_t(Q(X_q)) = \\
&&=\ln(\lambda) -\ln(\lambda(X_q))- P(X_q)\pa_t(Q(X_q))\ .
\eea
By comparison we conclude not only that $\pa_t\Omega = \ln(\lambda)$
but also
\be
\ln(\lambda(X_q)) =- P(X_q)\pa_t(Q(X_q))\ .
\ee
The proof of the Lemma is complete. Q.E.D.\par \vskip 4pt

\section{Planar Free energy for spectral curves of arbitrary genus}
\label{genuss}
The situation in the case the spectral curve is of higher genus is
only slightly different.
 We will  work with the following data:
a (smooth) curve $\Sigma_g$ of genus $g$ is assigned with two marked
points $\infty_Q$, $\infty_P$. On the curve we are given two functions
$P$ and $Q$ which have the following pole structure;
\begin{enumerate}
\item The function $Q$ has a simple pole at $\infty_Q$ and a pole of
  degree $d_2$ at $\infty_P$.
\item The function $P$ has a simple pole at $\infty_P$ and a pole of
  degree $d_1$ at $\infty_Q$.
\end{enumerate}
From these data it would follow that $P,Q$ satisfy a polynomial
relation but we will not need it for our computations.
By their definition we have 
\bea
P &=& \sum_{K=1}^{d_1+1} u_K Q^{K-1} -\frac t Q+ \mathcal O (Q^{-2}) =:
V_1'(Q) -\frac t Q+ \mathcal O (Q^{-2}) 
\ ,\ \ \hbox{near $\infty_Q$}\\
Q &=&\sum_{J=1}^{d_2+1} v_J P^{K-1} -\frac t P+ \mathcal O (Q^{-2})=: V_2'(P) -\frac t P+ \mathcal O (Q^{-2})
\ ,\ \ \hbox{near $\infty_P$}.
\eea
The fact that the coefficient of the power $Q^{-1}$ or $P^{-1}$ is the
same follows immediately from computing the sum of the residues of
$P{\rm d}Q$ (or $Q{\rm d}P$).
The formulas for the coefficients $u_K$, $v_J$, $t$ are the same as in
the genus zero case, viz
\be
u_K =-\res{\infty_Q} PQ^{-K}{\rm d}Q\ ,\ \ v_J
=-\res{\infty_P}QP^{-J}{\rm d}P\ ,\ \ t=\res{\infty_Q}P{\rm d}Q=
\res{\infty_P}Q{\rm d}P\ .
\ee
Note that the requirement that the curve possesses two meromorphic
functions with this pole structure imposes strong constraints on the
moduli of the curve itself. In fact a Riemann-Roch argument shows that
the moduli space of these data is of dimension $d_1+d_2+3+g$; so far
our data show only $(d_1+1)+(d_2+1)+1$ parameters and therefore we add
as parameters the following period integrals which, in the
matrix-model literature are referred to as {\em filling fractions}
\be
\epsilon_i := \frac 1{2i\pi}\oint_{a_i}P{\rm d}Q\ ,\ \ i=1,\dots g.
\ee
Here we have introduced a symplectic basis $\{a_i,b_i\}_{i=1\dots g}$ in the homology of the
curve $\Sigma_g$ and the choice of the $a$-cycles over the $b$-cycles
is purely conventional.\par
In this extended setting the free energy is defined by the equations
\bea
&&\pa_{u_K}\mathcal F_g = -\res{\infty_Q} PQ^K{\rm d}Q\ ,\qquad 
\pa_{v_J}\mathcal F_g = -\res{\infty_P} QP^J{\rm d}P\ ,\\
&& \pa_{\epsilon_i} \mathcal F_g = \oint_{b_j} P{\rm d}Q =:\Gamma_i.
\eea
Once more we introduce the exterior potentials by the same
requirements as in the genus $0$ case;
\bea
\Phi_1 =-\frac 1{2i\pi}\oint \ln\le(1-\frac {\tilde Q}Q\ri)\tilde P{\rm
  d}\tilde Q = -V_1(Q)+t\ln(Q) +\int_{X_q}P{\rm d}Q= \mathcal O(Q^{-1})\ ,\
\ \hbox{near $\infty_Q$}\\
\Phi_2 =-\frac 1{2i\pi}\oint \ln\le(1-\frac {\tilde P}P\ri)\tilde Q{\rm
  d}\tilde P = -V_2(P) + t\ln(P)+\int_{X_p}Q{\rm d}P = \mathcal O(P^{-1})\
,\ \ \hbox{near $\infty_P$}.
\eea
In the loop integral formulas the contours wind around the marked points so
as to leave the point where the potentials are computed inside, i.e.,
for instance, $\tilde Q/Q<< 1$.
As in the genus $0$ case the two points $X_p$, $X_q$ are implicitly
defined by the requirement $\mathcal O$(local parameter).
Recall the definition of the ``chemical'' potential
\be
\mu = Q(X_p)P(X_p)+\int_{X_p}^{X_q}\!\!\!\! P{\rm d}Q\ .
\ee
\bt
\label{freegenus}
The free energy over the curve $\Sigma_g$ is given by 
\be
2\mathcal F_g =  \res{\infty_Q}P\Phi_1{\rm d}Q + \res{\infty_P}
Q\Phi_2{\rm d}P + \frac 12 \res{\infty_Q}P^2Q{\rm d}Q + t\mu +
\sum_{i=1}^{g} \epsilon_i\Gamma_i\ .
\ee
\et
{\bf Proof}.
It is essentially the same as in the genus zero case. We start from
the same expression $\mathcal F_0$ used in the proof there (\ref{F0})
and proceed to the variation w.r.t. $u_K$. Following the same steps we
obtain the expression 
\bea
4i\pi\pa_K{\mathcal F}_0 \hspace{-18pt}&& =  4i\pi U_k +  \oint_{\infty_Q}\!\!\!\!\!
  P \le[\int_{X_q}\!\!\! (\pa_k P)_Q{\rm d}Q\ri] {\rm d}Q  + 
\oint_{\infty_P}\!\!\!\!\!  Q \le[\int_{X_q} \!\!\!(\pa_k Q)_P{\rm d}P\ri] {\rm d}P
+
\oint_{\infty_Q}\!\!\!\!\! QP {\rm d}\le[\int_{X_q}\!\!\! (\pa_k P)_Q{\rm d}Q\ri]+\\
&&
-2i\pi t\underbrace{\le[P(X_q)\pa_K(Q(X_q)) +Q(X_p)\pa_K (P(X_p)) -
  \int_{X_p}^{X_q}(\pa_KQ)_P{\rm d}P\ri]}_{\ds =\pa_K\mu}\ ,
\eea
where the subscripts in the contour integrals specify the point around
which we are circulating. The integration by parts of the third term
gives 
\bea
4i\pi\pa_K{\mathcal F}_0 \hspace{-18pt}&& =  4i\pi U_k +  \oint_{\infty_Q}\!\!\!\!\!
  P \le[\int_{X_q}\!\!\! (\pa_k P)_Q{\rm d}Q\ri] {\rm d}Q  + 
\oint_{\infty_P}\!\!\!\!\!  Q \le[\int_{X_q} \!\!\!(\pa_k Q)_P{\rm
    d}P\ri] {\rm d}P+\\
&& -
\oint_{\infty_Q}\!\!\!\!\! P\le[\int_{X_q}\!\!\! (\pa_k P)_Q{\rm
    d}Q\ri]{\rm d}Q-
\oint_{\infty_Q}\!\!\!\!\! Q\le[\int_{X_q}\!\!\! (\pa_k P)_Q{\rm
    d}Q\ri]{\rm d}P-2i\pi t\pa_K\mu=\\
&& = 4i\pi U_k +  \oint_{\infty_P}\!\!\!\!\!  Q \le[\int_{X_q} \!\!\!(\pa_k Q)_P{\rm
    d}P\ri] {\rm d}P + \oint_{\infty_Q}\!\!\!\!\! Q\le[\int_{X_q}\!\!\! (\pa_k Q)_P{\rm
    d}P\ri]{\rm d}P-2i\pi t\pa_K\mu \ ,
\eea
where we have used the thermodynamical identity (\ref{thermos}). Due
to the genus of the curve the two contours are not homologically
opposite and the sum of the two integrals gives finally (using the Riemann
bilinear identity which we recall in Appendix \ref{Riemann}) 
\bea
4i\pi\pa_K{\mathcal F}_0 \hspace{-18pt}&& = 4i\pi U_k - 2i\pi\sum_{i=1}^g \le[
\oint_{a_i} P{\rm d}Q \oint_{b_i} (\pa_KP)_Q{\rm d}Q - \oint_{b_i}
P{\rm d}Q \oint_{a_i} (\pa_KP)_Q{\rm d}Q \ri] - 2i\pi t\pa_K\mu = \\
&&=4i\pi U_k -2i\pi \sum_{i=1}^g \le[
\epsilon_i\pa_K\Gamma_i -
\Gamma_i\underbrace{\pa_K\epsilon_i}_{=0}\ri] - 2i\pi t\pa_K\mu =\\
&&=4i\pi U_k - 2i\pi\sum_{i=1}^g 
\epsilon_i\pa_K\Gamma_i    - 2i\pi t\pa_K\mu\ .
\eea
This implies promptly that 
\be
2\pa_K\mathcal F_g = 2\pa_K\mathcal F_0 + \sum_{i}
\epsilon_i\pa_K\Gamma_i + t\pa_K \mu = 2U_K\ ,
\ee 
as desired. The check for the variation w.r.t. the filling fractions
is the same by noticing that 
\bea
(\pa_{\epsilon_i}P)_Q = \mathcal O(Q^{-2})\ ;\ \ (\pa_{\epsilon_i}Q)_P
= \mathcal O(P^{-2})\ ,
\eea 
near the corresponding marked points. Moreover, by definition
\be
\oint_{a_j}(\pa_{\epsilon_j}P)_Q{\rm d}Q = \delta_{ij}\ .
\ee
 From this, following the same
formal steps and using the Riemann bilinear identity one easily finds 
\bea
2\pa_{\epsilon_i}\mathcal F_0 =  \Gamma_i- \sum_{j=1}^g 
\epsilon_j\pa_{\epsilon_i}\Gamma_j -  t\pa_{\epsilon_i}\mu\ ,\\ 
2\pa_{\epsilon_i}\mathcal F_g = 2\pa_{\epsilon_i}\mathcal F_0+
 t\pa_{\epsilon_i}\mu + \Gamma_i+  \sum_{j=1}^g\epsilon_j\pa_{\epsilon_i}\Gamma_j = 2\Gamma_i\ .
\eea
This concludes the proof. Q.E.D. \par\vskip 4pt
\br
In the two-matrix model setting the moduli of the spectral curve are
fixed uniquely in terms of the potentials $V_1,V_2$ by the requirement
$\Gamma_i=0,\ i=1,\dots,g$ (which is a minimum requirement)
\er
\br
The case of the one matrix model is a subcase of this setting where
one of the two potentials is quadratic, say $V_2$; in this case the spectral
curve is hyperelliptic of genus $[(d_1+2)/2]$.
\er
As for the previous case some special care must be paid for the
derivative w.r.t. $t$, but the result is the same as in the genus zero
case and it is contained in the next corollary.
\bc
The chemical potential $\mu$ is indeed the derivative $\pa_t\mathcal
F_g$. Moreover we have the formula
\be
\mu =  \res{\infty_Q} \big[V_1(Q) -
t\ln(Q/\lambda)\big]{\rm d}S - \res{\infty_P} \big[V_2(P) -
t\ln(P \lambda)\big]{\rm d}S  -\res{\infty_Q} PQ{\rm d}S +
\sum_{i=1}^g \epsilon_i\oint_{b_i} {\rm d}S\ ,\label{chemg}
\ee
where ${\rm d}S$ is the normalized differential of the third kind with
poles at $\infty_{P,Q}$ and residues $\pm 1$ and the function
$\lambda$ is the following function (defined up to a multiplicative
constant) on the universal covering of the
curve with a simple zero at $\infty_Q$ and a simple pole at $\infty_P$
\be
\lambda:= \exp\le(\int {\rm d}S\ri)\ . 
\ee
\ec
{\bf Proof.}
The proof is quite more involved than in genus zero  and requires some preparation.\par
We
introduce the normalized differential of the third kind with simple poles at
$\infty_{P,Q}$ and residues $\pm 1$,
\be
{\rm d}S = {\rm d}S_{\infty_Q,\infty_P}\ , \ \ \oint_{a_i}{\rm d}S =
0\ ,\ i=1,\dots g\ ;\ \  \res{\infty_Q}{\rm d}S = -1=-\res{\infty_P}{\rm
  d}S\ .\label{fff}
\ee
This provides a coordinate on the (covering of the) curve as follows 
\be
\lambda:= {\rm e}^{\int{\rm d}S}\ .
\ee
Quite clearly the parameter $\lambda$ is defined up to a
multiplicative constant and it is multiplicatively multivalued on the
curve $\Sigma_g$ (around the $b$-cycles). This arbitrariness and
multivaluedness will not affect our computations because formula
(\ref{fff}) remains unchanged under the rescaling $\lambda\mapsto
\delta\lambda$. Moreover it has no branch-points at $\infty_{P,Q}$,
where instead it has a simple zero and a simple pole (and no other zeroes or singularities). By
the definition then 
\be
{\rm d}S = {\rm d}\ln(\lambda)\ .
\ee
The rest mimics the genus zero case. We introduce the two
points $\Lambda_{p,q}$ such that 
\bea
&& \res{\infty_Q}\le(-V_1(Q) + t\ln(\lambda)+
\int_{\Lambda_q}\!\!\!\! P{\rm d}Q\ri){\rm d} S   = 0\ ,\\
&& \res{\infty_P}\le(-V_2(P) -t\ln(\lambda)+
  \int_{\Lambda_p}\!\!\!\! Q{\rm d}P\ri){\rm d}S  =0\ .
\eea
From the definition of the points $X_{p,q}$ and $\Lambda_{p,q}$ and following the same steps that were taken in
genus $0$ we have 
\bea
&& \int_{\Lambda_q}^{X_q} P{\rm d}Q = \res{\infty_Q} \le(V_1(Q) -
t\ln(Q/\lambda)\ri){\rm d}S \\
&&  \int_{\Lambda_p}^{X_p} Q{\rm d}P = -\res{\infty_P} \le(V_2(P) -
t\ln(P/\lambda)\ri){\rm d}S\ .
\eea
The next relation is different from the genus zero case:
\be
 P(\Lambda_q)Q(\Lambda_q) + \int_{\Lambda_q}^{\Lambda_p}\!\!\!\!Q{\rm
   d}P =- \res{\infty_Q} PQ{\rm d}S +
 \sum_{i=0}\epsilon_i\oint_{b_i}{\rm d}S\ .
\ee
Indeed we have
\bea
0\hspace{-18pt}&&= \oint_{\infty_Q}\le(\int_{\Lambda_q}\!\!\!P{\rm d}Q
- t\ln\lambda\ri){\rm d}S= \oint_{\infty_Q}
\le[QP - Q(\Lambda_q)P(\Lambda_q)
  -\int_{\Lambda_q}^{\Lambda_p}\!\!\!\!Q{\rm d}P - t\ln(\lambda)
  -\int_{\Lambda_p}\!\!\!\!Q{\rm d}P\ri]{\rm d}S=\\
&&=2i\pi\le(\res{\infty_Q}QP{\rm d}S\ri) +  2i\pi\le[Q(\Lambda_q)P(\Lambda_q)
  +\int_{\Lambda_q}^{\Lambda_p}\!\!\!\!Q{\rm d}P\ri] - \underbrace{\oint_{\infty_P}\le(\int_{\Lambda_p}\!\!\!Q{\rm d}P
- t\ln\lambda\ri){\rm d}S}_{=0} +
\sum_{i=1}^g\epsilon_i\oint_{b_i}{\rm d}S \ , 
\eea
where we have used the bilinear Riemann identity as well as the fact
that $\oint_{a_i}{\rm d}S=0$ and the fact that (by definition of the
point $\Lambda_p$) the term with the under-brace is residue-free at
$\infty_P$\footnote{Note that  $Q{\rm d}P-t{\rm d}S$ is a meromorphic
  Abelian differential without residues at the poles $\infty_{P,Q}$.}.
 We can now proceed as in 
Lemma \ref{lemma1} and obtain the formula
\bea
\mu \hspace{-18pt}&&= Q(X_p)P(X_p)+ \int_{X_p}^{X_q}\!\!\!\! P{\rm d}Q
=\nonumber \\
&&= \res{\infty_Q} \big[V_1(Q) -
t\ln(Q/\lambda)\big]{\rm d}S - \res{\infty_P} \big[V_2(P) -
t\ln(P \lambda)\big]{\rm d}S  -\res{\infty_Q} PQ{\rm d}S +
\sum_{i=1}^g \epsilon_i\oint_{b_i} {\rm d}S\ .\label{lemma1genus}
\eea
The formula (\ref{lemma1genus}) is the equivalent in higher genus of
the formula in Lemma \ref{lemma1}.\par
Coming back to the proof of the Corollary we note as in Corollary
\ref{Ft}  that  
\be
(\pa_t P)_Q{\rm d}Q = -(\pa_t Q)_P{\rm d}P= dS
\ee
is the {\em normalized} Abelian differential of the third kind with
poles at the marked points for we have 
\be
0 =\pa_t\epsilon_i = \oint_{a_i} (\pa_t P)_Q{\rm d}Q\ .
\ee
Computing the variation of $\mathcal F_0$ we now obtain 
\be
2\pa_t\mathcal F_0 =  - t\pa_t \le[ Q(X_p)P(X_p)+ \int_{X_p}^{X_q}\!\!\!\!  Q
{\rm d}P\ri] -\res{\infty_Q}PQ{\rm d}S+ \res{\infty_Q} 
  \ln\le(\frac Q \lambda\ri) P{\rm d}Q-\res{\infty_P}
  \ln(P\lambda)Q{\rm d}P
\ee
We now  have 
\bea
&&\res{\infty_Q} 
  \ln\le(\frac Q \lambda\ri) P{\rm d}Q = \res{\infty_Q}
  \le(V'_1(Q)-\frac tQ + \mathcal O(Q^{-2})\ri)  \ln\le(\frac Q
  \lambda\ri) {\rm d}Q=\\
&&= \res{\infty_Q}
  \le(V'_1(Q)+ \mathcal O(Q^{-2})\ri)  \ln\le(\frac Q
  \lambda\ri) {\rm d}Q-t \res{\infty_Q}\ln\le(\frac Q
  \lambda\ri) \frac{{\rm d}Q}Q=\\
&&= \res{\infty_Q}(V_1(Q){\rm d}S - t\ln(Q/\lambda){\rm d}S)\ ,
\eea
where, in the second residue, we can replace ${\rm d}Q/Q$ by ${\rm
  d}S$ because $\ln(Q/\lambda)=\mathcal O(1)$. A similar argument goes
for the term involving $P$ so that we finally have 
\be
2\pa_t\mathcal F_0 =  -t\pa_t \mu -\res{\infty_Q}PQ{\rm d}S+
\res{\infty_Q}\le(V_1(Q)-t\ln(Q/\lambda)\ri){\rm d}S
-\res{\infty_P}\le(V_2(P)-t\ln(P\lambda)\ri){\rm d}S\ .
\ee
Inserting this into the full expression of $\mathcal F_g$ we obtain 
\bea
2\pa_t\mathcal F_g \hspace{-18pt}&& = 2\pa_t\mathcal F_0 +\pa_t(t\mu) +\sum_{i=1}^g
\epsilon_i\pa_t\Gamma_i =\\
&&=  \mu -\res{\infty_Q}PQ{\rm d}S+
\res{\infty_Q}\le(V_1(Q)-t\ln(Q/\lambda)\ri){\rm d}S
-\res{\infty_P}\le(V_2(P)-t\ln(P\lambda)\ri){\rm d}S +\sum_{i=1}^g
\epsilon_i\oint_{b_i}{\rm d}S=\\
&&=2\mu\ ,
\eea
where we have used the expression (\ref{lemma1genus}).  This concludes
the proof of the corollary. Q.E.D. \par\vskip 4pt
\br
The formula for $\mu$ seems not symmetric in the r\^oles of $P$ and
$Q$ only superficially. In fact, if we exchanged the r\^oles, the
$3^{rd}$ kind differential should also change sign.
\er
We now investigate the scaling property of this Free energy. First off
we have the simple
\bl
\label{chemglemma}
Under the change of scale for the functions $Q = \delta \widetilde Q$ and $P=\sigma
\widetilde P$ the chemical potential rescales with an anomaly as
follows
\be
\mu = \delta\sigma \widetilde \mu + \delta\sigma \widetilde t \ln(\delta\sigma)\ .
\ee
\el
{\bf Proof.}
The proof is almost immediate from the expression (\ref{chemg})
considering the fact that under that rescaling we have
\be
u_K = \sigma \delta^{1-K}\widetilde{u_K}\ ;\ \ v_J =
\delta \sigma^{1-J}\widetilde{v_J}\ ;\ \ t=\delta\sigma \widetilde t\ ,\epsilon_i
= \delta\sigma \widetilde {\epsilon_i}\  .
\ee
On the other hand the differential ${\rm d}S$ (and hence the function
$\lambda$) are invariant as follows from its expression 
\be
{\rm d}S = (\pa_tP)_Q{\rm d}Q = \pa_{\widetilde t} \widetilde P{\rm
  d}\widetilde Q\ .
\ee
The proof follows then immediately from
(\ref{chemg}). Q. E. D.\par\vskip 4pt
With the above lemma we can immediately find the scaling properties of
the genus $g$ free energy
\bc[Scaling properties]
\label{scalg}
The Free energy of the genus $g$ data above satisfies the scaling
constraints
\bea
2\mathcal F_g = \le[\sum_K (1-K)u_K\pa_{u_K} + \sum_J v_J\pa_{v_J} +
  \sum_{i=1}^g\epsilon_i\pa_{\epsilon_i}+ t\pa_t \ri]\mathcal F_g -
\frac 12 t^2\ ,\label{scal1}\\
2\mathcal F_g = \le[\sum_K u_K\pa_{u_K} + \sum_J(1-J) v_J\pa_{v_J} +
  \sum_{i=1}^g\epsilon_i\pa_{\epsilon_i}+ t\pa_t \ri]\mathcal F_g -
\frac 12 t^2\ .\label{scal2}
\eea
\ec
{\bf Proof.} 
The proof is immediate from the definitions of the various objects and
using the anomalous scaling of the chemical potential $\mu$ in
Lemma \ref{chemglemma}. The constraint (\ref{scal1}) is obtained by
keeping $\sigma=1$ and differentiating w.r.t. $\delta$ at $\delta=1$,
and (\ref{scal2}) is obtained similarly by interchanging the r\^oles
of $\delta$ and $\sigma$ in the above procedure. Q.E.D.\par\vskip 4pt
The two scaling  constraints (\ref{scal1}, \ref{scal2}) form the ``two halves''
of the scaling 
\be
4\mathcal F_g =  \le[\sum_K (2-K)u_K\pa_{u_K} + \sum_J (2-J)v_J\pa_{v_J} +
 2 \sum_{i=1}^g\epsilon_i\pa_{\epsilon_i}+2 t\pa_t \ri]\mathcal F_g -
t^2\ ,
\ee
(obtained by adding (\ref{scal1}) and (\ref{scal2}))
which is the translation of the well--known property (\ref{scall}) for
higher genus Free energies. Of course the same properties
(\ref{scal1}, \ref{scal2}) hold also for the free energy in Section
\ref{freesec} (in which case the part involving the filling fraction
would be missing).\par\vskip 8pt 

We conclude with the remark that -quite clearly-  the (multivalued) function $\lambda$ is playing
essentially the same r\^ole of the uniformizing parameter in genus
zero. The quantities
\be
\ln(\gamma) := -\res{\infty_Q}\ln(Q/\lambda){\rm d}S\ ,\ \ \ln(\tilde \gamma):=
\res{\infty_P} \ln(P\lambda){\rm d}S\ ,
\ee
are the translation in this setting of the homonymous quantity in the
genus zero case, except that they need not be equal. However, since
now $\lambda = \exp\int{\rm d}S$ is defined up to a multiplicative
constant depending on the base--point of the integral, there would be
a choice for the base--point which makes $\gamma = \tilde
\gamma$. 
\section{Conclusion}
The formulas we have presented fill a gap in both the theory of the
dispersionless Toda hierarchy and the two-matrix model, where the
tau-function (free energy in the planar limit) is known only through
its partial derivatives but no closed formula for the tau-function
itself is known.\par
The derivation and the technique of the proof emphasizes the
importance of the spectral curve of the model, at least in the case of
polynomial potentials or, in the dToda language, finite Laurent
polynomials for the Lax operators.\par
On a slightly different perspective, we have computed the free energy
of the matrix model in the case where the spectral curve is of genus
$g>0$; the computation is less explicit than in genus zero but the
formula is closed. \par
It would be interesting to explore further the remnant of the
Poisson structure (\ref{dToda}) in this context. In fact it is almost
immediate to verify that still
\be
\{P,Q\} = \lambda\pa_\lambda P \pa_t Q-\lambda\pa_\lambda Q\pa_tP = 1 .
\ee 
The investigation of the Poisson structure  will be the
topic of subsequent publications. 
\par\vskip 5pt

{\bf Acknowledgements}\\
I would like to thank Bertrand Eynard for discussion and pointing out
many relevant references, and Dmitri Korotkin for help in the
computation of the higher genus case. 
Also I would like to thank the anonymous referee who helped me realize a mistake
in a first version, which prompted  me towards formulation of Corollary \ref{scalg}.\\
Many thanks go also to my daughter Olenka for anticipated inspiration.

\appendix
\section{The Riemann bilinear identity}
\label{Riemann}
It is a classical identity but it may be useful to recall it here.
Let be given a curve $\Sigma_g$ of genus $g$ and a symplectic basis in the
homology $\{a_i, b_i\}_{i=1\dots g}$. Let $\eta$ and $\omega$ be
two meromorphic Abelian differentials and $\omega$ be without residues and define 
\be
\Omega=\int_P\omega\ .
\ee 
Let $\widetilde \Sigma_g$ be a simply connected fundamental domain on the
universal covering of the curve then the  function
$\Omega$ is single-valued inside  $\tilde \Sigma_g$ with possibly
poles.
Let $\pa\widetilde\Sigma_g$ be the boundary of the domain constituted
by the cycles of the chosen basis. Then the bilinear Riemann identity
claims
\be
2i\pi\!\!\!\!\!\sum_{\hbox{residues}} \!\!\!\!\!\eta\Omega = \sum_{i=1}^g\le[
\oint_{a_i}\eta\oint_{b_i}\omega -
\oint_{a_i}\omega\oint_{b_i}\eta\ri]\ .
\ee
The proof is very easy and can be found in \cite{DNF}

\end{document}